\documentclass{article}
\usepackage{spconf,amsmath,amsfonts,epsfig,subfigure}

\newtheorem{thm}{Theorem}

\newtheorem{cor}{Corollary}

\def \R { {\mathbb R} }

\title{ADAPTIVE COMPRESSED SENSING -- A NEW CLASS OF SELF-ORGANIZING CODING MODELS FOR NEUROSCIENCE}
\name{ William K. Coulter*, Christopher J. Hillar**\thanks{Hillar is supported under an NSA Young Investigators Grant and an MSRI postdoctoral fellowship}, Guy Isley*, Friedrich T. Sommer*}
\address{*: University of California, Berkeley, 156 Stanley Hall, MC\# 3220, Berkeley, CA 94720-3220\\
     **: Mathematical Sciences Research Institute, 17 Gauss Way, Berkeley, CA 94120 \\
     \texttt{\ninept wcoulter@berkeley.edu, chillar@msri.org, gisley@gmail.com, fsommer@berkeley.edu}}

\begin{document}
\ninept
\maketitle
\begin{abstract}
Sparse coding networks, which utilize unsupervised learning to maximize coding efficiency, have successfully reproduced response properties found in primary visual cortex \cite{AN:OlshausenField96}.  However, conventional sparse coding models require that the coding circuit can fully sample the sensory data in a one-to-one fashion, a requirement not supported by experimental data from the thalamo-cortical projection. To relieve these strict wiring requirements, we propose a sparse coding network constructed by introducing synaptic learning in the framework of compressed sensing.  
We demonstrate a new model that evolves biologically realistic, spatially smooth receptive fields despite the fact that the feedforward connectivity subsamples the input and thus the learning must rely on an impoverished and distorted account of the original visual data. Further, we demonstrate that the model could form a general scheme of cortical communication: it can form meaningful representations in a secondary sensory area, which receives input from the primary sensory area through a ``compressing'' cortico-cortical projection.  Finally, we prove that our model belongs to a new class of sparse coding algorithms in which recurrent connections are essential in forming the spatial receptive fields.  
\end{abstract}

\section{Introduction}
\vspace{-5pt}

Guided by the early ideas on efficient sensory coding \cite{AN:Attneave54,AN:Barlow83}, 
self-organizing network models for sparse coding have been critical in understanding how essential response properties, such as orientation selectivity, are formed in sensory areas through development and experience \cite{AN:OlshausenField96,AN:BellSejnowski97}. There is now a wealth of such models, all based on a set of similar connectivity patterns: a neuron receives feedforward drive from the afferent input and competes with other neurons in the network through mainly inhibitory lateral connections, see \cite{AN:Foeldiak90,AN:OlshausenField96,AN:RehnSommer07,AN:Rozelletal08,AN:YinEtal08}. Some of these models are capable of reproducing the response properties in primary visual cortex quantitatively, for instance, the network model proposed in \cite{AN:RehnSommer07} that implements an algorithm called optimized orthogonal matching pursuit \cite{AN:Rebollo-NeiraLowe02}. 

While these models match physiological data quite impressively, their correspondence to the anatomical connectivity in cortex is problematic. According to the models, 
the neurons must have access to the full data, for instance, to all pixels of an image patch. Many models even suggest that each neuron has the feedforward wiring in place so that the synaptic structure in the feedforward path can match the receptive field exactly. It is unclear if the development of the thalamic projections into V1 can reach such connection density and microscopic precision  -- even though thalamic receptive fields do match the receptive fields of monosynaptically connected V1 cells with some precision \cite{AN:Chapmanetal91,AN:ReidAlonso95}. Here, we explore learning schemes for neural representations that relieve these requirements on the feedforward wiring. In addition, we assess the ability of these learning schemes to account for learning in cortico-cortical projections, for which it has been established that only a fraction of local cells in the origin area send fibers to a target area \cite{AN:SchuezEtal06}. Therefore, conventional sparse coding at the receiver end can not work. 

To construct sparse coding networks with less restrictive wiring conditions we build on {\it compressed sensing} or {\it compressed sampling}, a method originally developed for data compression by subsampling. The decompression step in these algorithms has a close similarity to sparse coding models and thus these methods can form a framework for developing a new class of neural networks for self-organizing neural representations in cortical areas. Specifically, we explore the hypothesis of a generic scheme of cortical communication in which each cortical area unwraps subsampled input data into a sparse code to perform local computations and then sends a subsampled version of its local representation to other cortical areas.

\section{Adaptive compressed sensing}
\vspace{-5pt}
\label{models}

\begin{figure*}[t]
\begin{center}
\subfigure[Feedforward weights and receptive fields of sparse coding circuit \cite{AN:RehnSommer07}. The patterns look indistinguishable to the eye and can be proven to be the same (see Thm~1).]{
\includegraphics[height=4cm, trim= 1.5in 4.9cm 1.5in 3.5cm, clip=true]{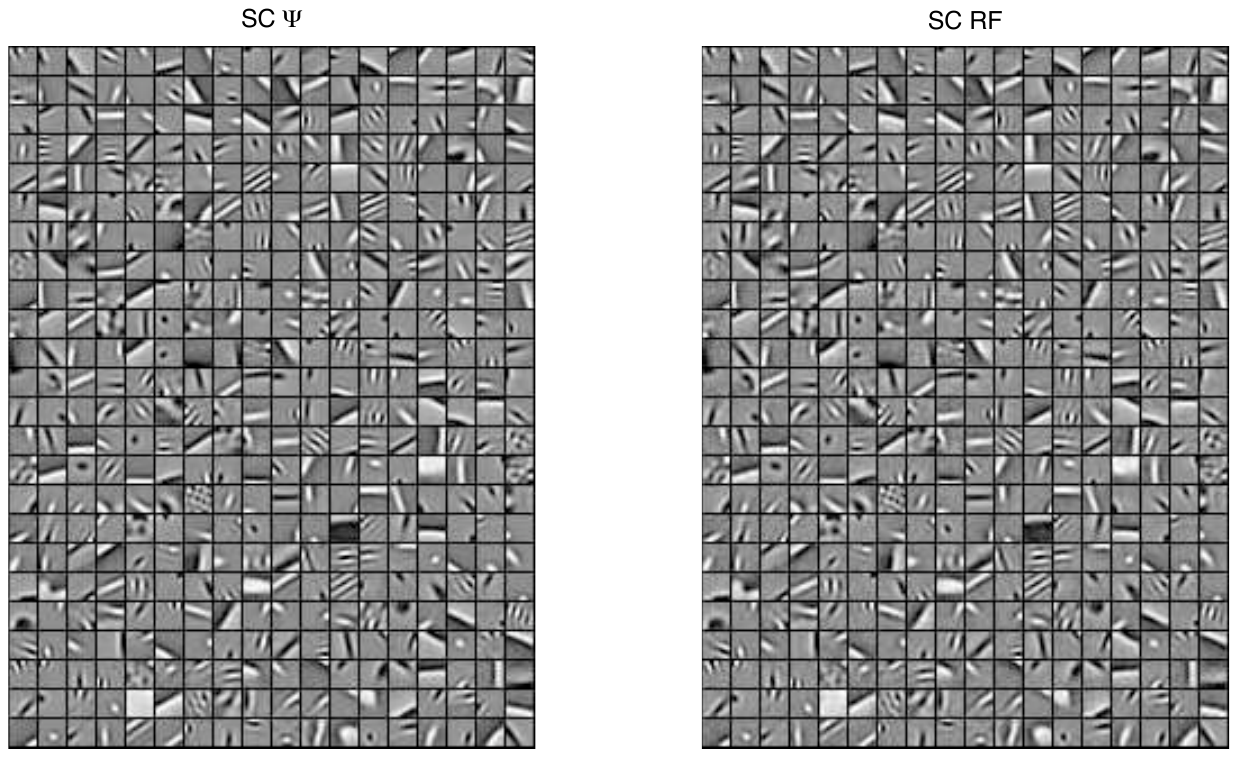}
\label{fig1a}
}
\hspace{5pt}
\subfigure[Feedforward weights (ACS FF) and receptive fields (ACS RF) of adaptive compressed sensing circuit. The plot ``ACS (2nd stage) RF'' depicts receptive fields learned in a cascaded secondary sensory area receiving $\bf \Phi_2 a(x)$ as the input.]{
\includegraphics[height=4cm, trim= 1.5in 7.2cm 1.5in 5.9cm, clip=true]{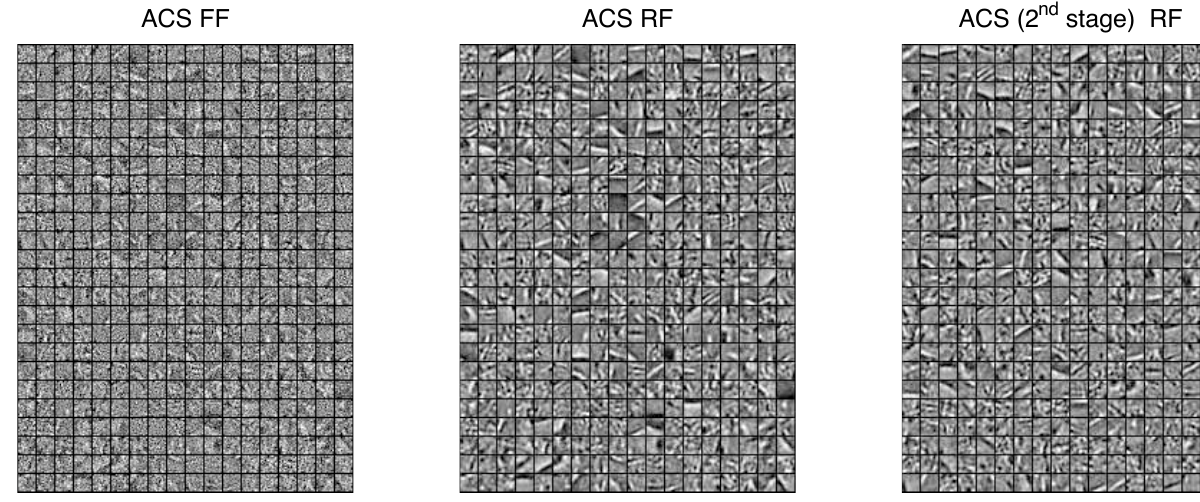}
\label{fig1b}
}
\vspace{-10pt}
\caption{Comparison of Feedforward weights (FF) and Receptive Fields (RF) of various models.}
\vspace{-20pt}
\end{center}
\label{fig1}
\end{figure*}
{\it Conventional sparse coding} is governed by the objective function
\begin{equation}
\setlength{\itemsep}{1pt}
\setlength{\parsep}{1pt}
\setlength{\parskip}{1pt}
E({\bf x, a, \Psi}) = \frac{1}{2}||{\bf x} - {\bf \Psi} {\bf a}||^2 + S({\bf a}).
\label{esparsecode}
\end{equation}
Here ${\bf x} \in \R^m$ is the input data, $\bf \Psi$ is a real $m \times n$ matrix whose columns form a dictionary for constructing the input, and ${\bf a}$ is a coefficient vector for this reconstruction: $\tilde{\bf x} = {\bf \Psi} {\bf a}$. The function $S({\bf a})$ is a sparseness constraint that penalizes neural activity and forces the coefficient vector to be sparse.

For a given input ${\bf x}$, the sparse coding operation is given by an energy minimization
\begin{equation}
{\bf a(x)} := \text{\rm arg}\min_a E({\bf x, a, \Psi}) \in \R^n.
\label{coding}
\end{equation}
We adapt the dictionary to the data by minimizing $E({\bf x, a(x), \Psi})$ from Eq.~\ref{esparsecode} and Eq.~\ref{coding} with respect to $\bf \Psi$. Using gradient descent for the adaptation yields a Hebbian synaptic learning rule for the $\bf \Psi$ components \cite{AN:OlshausenField96,AN:RehnSommer07}.

{\it Compressed sensing} is a technique for data compression using a random projection matrix $\bf \Phi$ to compress the data ${\bf x}  \in \R^m$ to ${\bf \Phi x}  \in \R^k$ with $k < m$. The decompression uses energy minimization (\ref{coding}) of an error-based energy function similar to Eq.~\ref{esparsecode}:
\begin{equation}
E({\bf x, a, \Phi}) = \frac{1}{2}||{\bf \Phi x} - {\bf \Phi \Psi} {\bf a}||^2 + S({\bf a}).
\label{ecs}
\end{equation}
The original data is reconstructed as $\tilde{\bf x} = {\bf \Psi \; a(\Phi x)}$. 
In conventional compressed sensing, a fixed dictionary $\bf \Psi$ is chosen.  The decompression can be shown to work if (i) a dictionary $\bf \Psi$ is used in which the data can be sparsely represented, (ii) matrices $\bf \Phi$ and $\bf \Psi$ are incoherent, and (iii) the dimension of data compression $k$ is larger than the sparsity of the data \cite{AN:CandesRombergTao06,AN:CandesWakin08}.

Building on a model by Rehn and Sommer \cite{AN:RehnSommer07}, we introduce {\it adaptive compressed sensing (ACS)}, an adaptive version of compressed sensing governed by:
\begin{eqnarray}
\label{acs}
E({\bf x, a, \Phi, \Theta}) = \frac{1}{2}||{\bf \Phi x} - {\bf \Theta} {\bf a}||^2   + \lambda||{\bf a}||_{L_0}\\ 
= - {\bf x}^{\top} {\bf \Phi^{\top} \Theta} {\bf a} + \frac{1}{2}{\bf a}^{\top} {\bf \Theta}^{\top}{\bf \Theta} {\bf a} + \lambda||{\bf a}||_{L_0}+ const \nonumber .
\end{eqnarray}
Learning is executed by gradient descent on $\bf \Theta$ in exactly the same fashion as in conventional sparse coding, e.g. \cite{AN:OlshausenField96,AN:RehnSommer07}. Note, however, the difference between ACS and conventional sparse coding. The new algorithm (\ref{acs}) forms a dictionary of the compressed data, the $k \times n$ matrix $\bf \Theta$, whereas conventional sparse coding forms a dictionary of the original data, an $m \times n$ matrix. 
Although we use the L$_0$-sparseness constraint to penalize the number of active units, $S({\bf a}) = \lambda ||{\bf a}||_{L0}$, similar schemes of adaptive compressed sensing can be realized with other types of sparseness constraints.

{\it Network implementation of ACS:} Analogous to earlier models of sparse coding, coding in ACS can be implemented in a network where each neuron $i$ computes the gradient of the two differentiable terms in Eq.~\ref{acs} as
\begin{equation}
\frac{\partial E'}{\partial a_i} = - ({\bf x}^{\top} {\bf \Phi^{\top} \Theta})_i + ({\bf \Theta}^{\top}{\bf \Theta} {\bf a})_i ,
\label{gradesparsecode}
\end{equation}
see \cite{AN:RehnSommer07} for further detail.
In the neural network for the ACS method the {\it feedforward weights} are $FF:= {\bf \Phi^{\top} \Theta}$ and the competitive {\it feedback weights} are $FB: = - {\bf \Theta}^{\top}{\bf \Theta}$. Note that if $\bf \Phi$ is the identity matrix, ACS coincides with conventional sparse coding for which the corresponding neural network would be defined by $FF = \bf \Psi$ and $FB = -FF^{\top} FF$ \cite{AN:OlshausenField96,AN:RehnSommer07}. The important difference between the two wiring schemes is that the feedforward weights of ACS subsample and mix the original data. Thus, coding and weight adapatation in ACS lack the full access to the original data that is available to conventional sparse coding. 
Remarkably, the simulation experiments described in the next section demonstrate that the neurons in the ACS network still develop biologically realistic receptive fields, despite the limited exposure to the original data.

\section{Simulation experiments with adaptive compressed sensing}
\vspace{-5pt}
\label{exp1}
\begin{figure*}[bt]
\begin{center}
\includegraphics[height=3cm, trim= 4cm 9.3cm 3.1cm 8.1cm, clip=true]{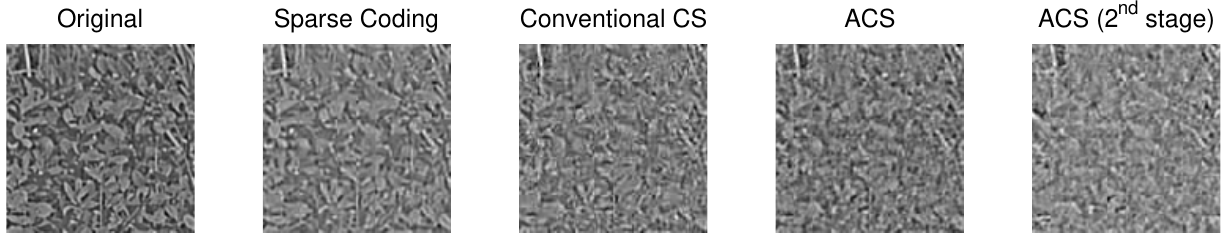}
\vspace{-15pt}
\caption{Original image and reconstruction using the different methods on 12 by 12 image patches. For reconstructing the images from the representations formed by ACS (in the first and 2nd stage), we used the receptive fields.}
\vspace{-25pt}
\label{fig2}
\end{center}
\end{figure*}
We compared the ability of networks described in section~\ref{models} to code patches of natural scene images and form receptive fields. The images were preprocessed by ``whitening,'' as described in \cite{AN:OlshausenField96}. The coding circuits encoded patches of $12\times 12$ pixels, making the dimension of the data $m=144$. For ACS, we used a sampling matrix $\bf \Phi$ that downsampled the original data to $k=60$ dimensions. All coding circuits contained $n=432$ neurons, thereby producing representations of the original data $a \in \R^n$ that were three times overcomplete. In addition to image coding in a primary sensory area, we also tested whether the ACS model could be used by a secondary sensory area (2nd stage). Our model of the 2nd stage receives a subsampled version of the sparse code generated in the primary visual area ${\bf \Phi_2 a(x)} \in \R^k$ and produces a sparse code ${\bf a_2} \in \R^n$, again with $k=60$ and $n=432$. Models used a coefficient $\lambda = 0.1$ in the sparseness constraint of Eq.~\ref{acs}.

Since the ACS model learns a dictionary of the compressed data rather than the original data, the original image cannot be 
reconstructed from the adapted $\bf \Theta$ matrix. Note that computing the data dictionary from $\bf \Theta$ requires an ill-posed step of matrix factorization: $\bf \Theta = \Phi \Psi$. Therefore, to assess the quality of the emerging codes in the ACS model, we measured receptive fields in the trained circuit (as physiologists do from the responses of real neurons). We compute the \textit{receptive fields} for a set $I$ of visual stimuli ${\bf x} \in \R ^m$ as
\begin{equation}
RF :=  \frac{1}{|I|} \sum_{{\bf x} \in I} {\bf x} \cdot {\bf a(x)}^{\top}.
\end{equation}
Notice that $RF$ is an $m \times n$ real matrix, the $i$-th column representing the receptive field of the $i$-th neuron.

Figure~\ref{fig1} shows the feedforward weights and the receptive fields of the different coding circuits. While the feedforward weights and receptive fields in Fig~\ref{fig1a} are very similar for sparse coding, they are markedly different for ACS in Fig~\ref{fig1b}. 
Interestingly, while subsampling makes the feedforward weights somewhat amorphous and noisy, the resulting receptive fields of ACS are smooth and resemble the receptive fields of sparse coding. When used in a secondary sensory area (2nd stage), ACS forms response properties that are similar to those in the primary sensory area, though the response properties differ on a neuron-by-neuron basis. 

To assess how well the sparse codes describe the original input, we computed image reconstructions. For sparse coding, we used the basis functions $\bf \Psi$; for ACS, the receptive fields $RF$. Fig~\ref{fig2} shows that ACS forms representations in the primary and secondary area that can be used for reconstruction, although the quality of reconstructions obtained from conventional sparse coding is not achieved. 

\begin{figure}[bh]
\vspace{-17pt}
\raggedleft
\includegraphics[width=0.8\columnwidth,trim= 2.5cm 1.3cm 2.9cm 0.5cm, clip=true]{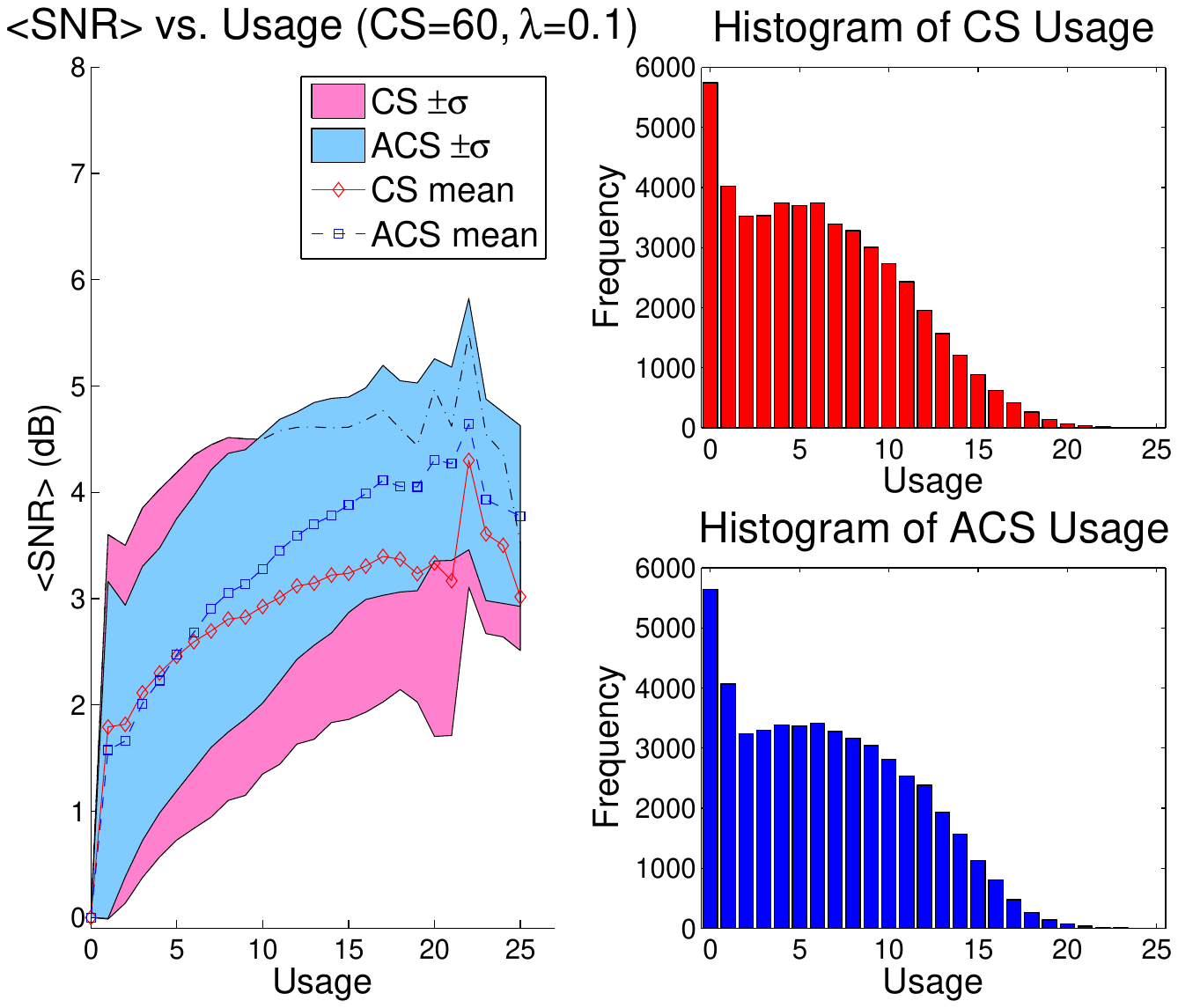}
\vspace{-15pt}
\caption{Left: Signal-to-noise ratio (mean and range of standard deviation) in reconstructions with conventional compressed sensing (red) and ACS (blue). Means do not differ significantly, but the performance of ACS has smaller variance. Right: Histograms of the number of neurons used per reconstruction are similar for both methods.}
\vspace{-22pt}
\label{fig3}
\end{figure}

Fig~\ref{fig3} compares the reconstruction quality of conventional compressed sensing (using the basis functions that were adapted to the original data) and adaptive compressed sensing. The mean reconstruction qualities do not differ, though ACS performs with lower variance over the set of input patches we tested.

These simulation results suggest that ACS is able to form representations of sensory data that convey its essential structure although the coding network receives only a subsampled version of the data. In the next section, we investigate the mathematical differences between conventional sparse coding and adaptive compressed sensing.
\section{Differences between ACS and conventional sparse coding models}
\vspace{-5pt}
In this section we derive two theorems to establish that ACS defines a class of sparse coding algorithms whose properties differ qualitatively from those of conventional sparse coding models. 
\vspace{-10pt}
\subsection{Receptive fields and feedforward weights coincide in conventional sparse coding networks}
Assuming that $\bf x$ is a column vector of random variables on a measure space $\Omega$ with probability measure $\mu$, the matrix $RF$ will be an approximation to the correlation of $\bf x$ and $\bf a(x)$ (which is assumed to have zero mean): $\text{Cor}({\bf x,a}) = \int {\bf x a(x)}^{\top} d \mu.$

The strong law of large numbers guarantees that given enough samples, the matrix $RF$ will be close to the integral above.  
For this reason, we will assume that $RF =  \text{Cor}({\bf x,a})$.  We are interested in calculating the necessary relationships between the quantities $RF$, $FF$, $FB$, $\bf \Phi$, and $\bf \Theta$. (Recall: $FF = \bf \Phi^{\top} \bf \Theta$ and $FB = -\bf \Theta^{\top} \bf \Theta$.)

In our setup, the data $\bf x$ are assumed to come from a sparse number $k$ of independent causes (nonzero values in $\bf a$).  Moreover, the method of recovering $\bf a(x)$ from a particular $\bf x$ is assumed to be exact (or near exact) in solving Eq.~\ref{coding} and \textit{independently distributed}; that is, $\bf \Phi x = \Theta \; a(x)$ and we have: $D = \int {\bf a(x)a(x)}^{\top} d\mu,$ in which $D$ is an $n \times n$ diagonal matrix.   
One now calculates:
\begin{equation}\label{phiRFeq}
{\bf \Phi} RF =  \int {\bf \Phi x \; a(x)}^{\top} d \mu =   {\bf \Theta} \int  {\bf a(x) a(x)}^{\top} d \mu =   {\bf \Theta} D.
\end{equation}
In particular, this implies the following.


\begin{thm}
If $\bf \Phi$ is the identity, then the receptive fields are scalar multiples of the feedforward weights.
\end{thm}



\subsection{Feedback co-shapes receptive fields in the ACS model}

In the compressive sensing regime, the matrix $\bf \Phi$ is no longer the identity but instead a compressive sampling matrix.  In this case, the receptive fields are almost never scalar multiples of the feedforward weights.  A precise analytic relationship is given by the following theorem.  As an important consequence, we obtain the  qualitative interpretation found in Theorem \ref{CorToThm3} below.  We omit here the proofs.

\begin{thm}\label{minRF-FF-thm}
If $FF$, ${\bf \Phi}$, and ${\bf \Theta}$ are nonzero and ${\bf \Theta}{\bf \Theta}^{\top}$ is invertible, then with $C =  \frac{|\text{\rm tr}(RF^{\top} FF)|}{||FF||^2 \cdot ||{\bf \Phi}|| \cdot ||{\bf \Theta}^{\top} ({\bf \Theta}{\bf \Theta}^{\top})^{-1}||}$, we have:
\[ \min_t ||RF - t FF|| \geq C \cdot \min_t ||tI - {\bf \Phi} {\bf \Phi}^{\top}||,\]
\end{thm}


What is important here is not the technical statement of Theorem \ref{minRF-FF-thm}, but rather the following qualitative versions.

\begin{cor}\label{mainFBCor}
If ${\bf \Phi}{\bf \Phi}^{\top}$ is not (close to) a scalar multiple of the identity, then $RF$ is not (close to) a scalar multiple of $FF$.
\end{cor}

\begin{thm}\label{mainFF-FBthm}
If the feedback weights are not a scalar multiple of $FF^{\top} FF$, then RF is not a scalar multiple of FF.
\end{thm}


Finally, we remark that in the compressive sensing regime $k \ll n$ and ${\bf \Phi}$ is a random matrix; thus, the hypothesis of the previous results are satisfied generically:

\begin{thm}\label{CorToThm3}
In adaptive compressed sensing, the receptive fields are almost surely not scalar multiples of the feedforward weights.
\end{thm}

\section{Discussion and Conclusions}
\vspace{-5pt}

We have proposed {\it adaptive compressed sensing (ACS)}, a new scheme of learning under compressed sensing that forms a dictionary adapted to represent the compressed data optimally. The coding and learning scheme of ACS can be formulated as a neural network, building on an earlier sparse coding model \cite{AN:RehnSommer07}.  Our model learns in the weights of the coding circuit while keeping the random projection fixed, as opposed to a previous suggestion which optimizes the compression performance by learning in the random projection \cite{AN:Weissetal07}. 

Our study focuses on the application of ACS to understand how cortical regions in ascending sensory pathways can analyze and represent signals they receive through thalamo-cortical or cortico-cortical connections. Conventional sparse coding theories were succcessful in reproducing physiological responses in primary sensory regions but they require exact matches between feedforward connections and receptive field patterns of cortical neurons  (see Theorem 1 and Fig~\ref{fig1a} for an example). Although it has been shown that thalamocortical wiring is to some extent specific \cite{AN:Chapmanetal91,AN:ReidAlonso95}, exact matches between feedforward circuitry and receptive fields are not supported by experimental data. In addition, a recent quantitative study of cortico-cortical projections suggests that the number of fibers reaching a target area can only be a fraction of the local neurons in the area of origin \cite{AN:SchuezEtal06}.   

We have tested if ACS could serve as a computational model for how cortical areas can form a representation of data received through afferent projections that subsample the activity pattern in the previous stage. We demonstrate that ACS can form representation of visual data, though, unlike in conventional sparse coding models, the coding circuit receives only a subsampled version of the original data. Further, we have demonstrated that the algorithm is stackable in a hierarchy. The sparse code formed by ACS in a primary sensory area, when sent through another compressing projection can be decoded in a secondary sensory area into another meaningful visual representation. The simulation results proof the concept that the ACS model can serve as a generic building block in a communication scheme  between cortical areas. The scheme consists of repeated cycles of compression and expansion. Specifically, a sparse local representations is compressed, sent through cortico-cortical projections and expanded to sparse local representations at the receiver end, reminiscent to Braitenberg's idea of the pump of thought \cite{AN:Braitenberg77}. The scheme of ACS suggests that representations in the brain can be sparse \cite{AN:DeWeeseZador06,AN:HromadkaEtal08} and dense \cite{ShadlenNewsome94,ShadlenNewsome95}, with the type of code being lamina-specific.  Regarding the still debated role of recurrent circuitry in producing orientation selectivity (e.g., \cite{AN:HubelWiesel62,AN:HirschEtal98,AM:FersterMiller00}), ACS suggests that if the input subsamples the data then feedback in shaping the receptive fields becomes essential for coding efficiency.

\bibliographystyle{IEEEbib}
\bibliography{strings,refs}

\end{document}